\documentclass[journal]{IEEEtran}
\usepackage{times}
\usepackage[final]{graphicx}
\usepackage{amsmath,amsfonts}
\usepackage{amssymb,amsbsy}
\usepackage{times}

\usepackage{psfrag}
\usepackage{epsf}
\usepackage{textcomp}
\usepackage{subfigure}

\IEEEoverridecommandlockouts  

\begin{document} 
\title{A Random Search Framework for Convergence Analysis of Distributed Beamforming with Feedback} 
\author{Che Lin, \emph{Member, IEEE}, Venugopal V. Veeravalli, \emph{Fellow, IEEE}, and Sean Meyn, \emph{Fellow, IEEE}
\thanks{This research was supported in part by the NSF awards \#CCF 0431088 and \#CNS 0831670, and ITMANET  DARPA \#RK 2006-07284 through the University of Illinois, 
and by a Vodafone Foundation Graduate Fellowship. Any opinions, findings, and conclusions or recommendations expressed in this material are those of the authors and do not necessarily reflect the views of NSF or DARPA.} \\[1mm] 
\thanks{Che Lin is with Institute of Communication Engineering, National Tsing Hua University, Hsinchu, 30013, Taiwan (e-mail: clin@ee.nthu.edu.tw).}
\thanks{Venugopal V. Veeravalli and Sean Meyn are with Coordinated Science Laboratory, University of Illinois, Urbana-Champaign, Urbana, IL 61801, USA (e-mail: \{vvv, meyn\}@illinois.edu).}
}

\maketitle

\newcommand\ignore[1]{}
\newtheorem{thm}{Theorem}
\newtheorem{lem}{Lemma}
\newtheorem{example}{Example}
\newtheorem{defn}{Definition}
\newtheorem{prop}{Proposition}
\newtheorem{remark}{Remark}
\newtheorem{assumption}{Assumption}
\newtheorem{property}{Property}
\newtheorem{conjecture}{Conjecture}
\newtheorem{algm}{Algorithm}
\newtheorem{problem}{Problem}

\newcommand{\note}{{\bf Note: }}
\newcommand{\beq}{\begin{equation}}
\newcommand{\eeq}{\end{equation}}
\newcommand{\beqa}{\begin{eqnarray}}
\newcommand{\eeqa}{\end{eqnarray}}
\newcommand{\mb}{\mathbf}
\newcommand{\mt}{\textrm}
\newcommand{\bnum}{\begin{enumerate}}
\newcommand{\enum}{\end{enumerate}}
\newcommand{\bu}{ {\bf u}} 

\newcommand{\tH}{\widetilde{{ \rm H}\phi}}
\newcommand{\tn}{\widetilde{n}}
\renewcommand{\th}{\widetilde{h}}
\newcommand{\tA}{\widetilde{{A}}}
\newcommand{\tY}{\widetilde{{Y}}}
\newcommand{\tW}{\widetilde{{W}}}
\newcommand{\tmH}{\widetilde{ \mb{H}}}
\newcommand{\tmW}{\widetilde{\mb{W}}}
\newcommand{\tmY}{\widetilde{\mb{Y}}}

\newcommand{\E}{\mathbb{E}}
\newcommand{\R}{\mathbb{R}}
\newcommand{\C}{\mathbb{C}}
\newcommand{\N}{\mathbb{N}}
\newcommand{\HH}{\mathbb{H}}
\renewcommand{\S}{\mathbb{S}}
\newcommand{\bQ} { {\bf Q} }

\newcommand{\sfh}{ {\sf{H}}}
\newcommand{\bmfm}{ {\sf{bf}}}
\newcommand{\bo} { {\mathbf{0}} }
\newcommand{\trace}{ {\mathrm{Tr}} }
\newcommand{\diag}{ \mathrm{diag} }

\newcommand{\Y}{\tilde{\mb{Y}}}
\newcommand{\bA} { {\mathbf{A}} }
\renewcommand{\H}{\tilde{\mb{H}}}
\renewcommand{\N}{\tilde{\mb{N}}}
\newcommand{\A}{\widetilde{A}}
\newcommand{\Ad}{\widetilde{A}\widetilde{A}^H}
\renewcommand{\b}{\mb{b}}

\newcommand{\pf}{{\bf Proof: }}

\newcommand{\bmY}  {{\mathbf{Y}} }
\newcommand{\bmX}  {{\mathbf{X}} }
\newcommand{\bmH}  {{\mathbf{H}} }
\newcommand{\bmW}  {{\mathbf{W}} }
\newcommand{\bmN}  {{\mathbf{N}} }

\newcommand{\bmx}  {{\mathbf{x}} } 
\newcommand{\bms}   {{\mathbf{s}} }
\newcommand{\bmn}   {{\mathbf{n}} }
\newcommand{\bmy}   {{\mathbf{y}} }
\newcommand{\bmh}   {{\mathbf{h}} }
\newcommand{\bmv}   {{\mathbf{v}} } 
\newcommand{\bma}   {{\mathbf{a}} } 
\newcommand{\snr}   {{\sf{SNR}}} 

\newcommand{\bLambda}  {{\mathrm{\Lambda}} } 
\newcommand{\bfLambda}  {{\mathbf{\Lambda}} } 
\newcommand{\bU}  {{\mathbf{U}} }
\newcommand{\bH}  {{\mathbf{H}} }
\newcommand{\bEe}{{\mathbf{E}}}    
\newcommand{\iid} {  {\sf{iid}} }
\newcommand{\ind} { {\sf{ind}} }

\newcommand{\mse}{ {\sf mse} }
\newcommand{\mmse}{ {\sf{mmse}} }
\newcommand{\bflambda}{ {\mathbf{\Lambda}} }
\newcommand{\bV} { {\mathbf{V}} }
\newcommand{\bv} { {\mathbf{v}} }
\newcommand{\bx} { {\mathbf{x}} }
\newcommand{\bP} { {\mathbf{P}} }
\newcommand{\bs} { {\mathbf{s}} }
\newcommand{\bS} { {\mathbf{S}} }
\newcommand{\ba} { {\mathbf{a}} }
\newcommand{\bb} { {\mathbf{b}} }
\newcommand{\bz} { {\mathbf{z}} }
\newcommand{\bZ} { {\mathbf{Z}} }
\newcommand{\bAt} { {\widetilde{\bA }} }
\newcommand{\stat} { {\sf stat} }
\newcommand{\bI}{{\bf I}}
\newcommand{\dg}{\dagger}

\newcommand{\vsp}{\vspace{0.1in} }
\newcommand{\hsp}{\hspace{0.1in} }
\newcommand{\vspp}{\vspace{0.05in} }
\newcommand{\hspp}{\hspace{0.05in} }
\newcommand{\hsppp}{\hspace{0.02in} }
\newcommand{\vsppp}{\vspace{0.02in} }

\newcommand{\vspn}{\vspace{-0.1in} }
\newcommand{\vspnn}{\vspace{-0.05in} }
\newcommand{\hspn}{\hspace{-0.1in} }
\newcommand{\hspnn}{\hspace{-0.05in} }

\newcommand{\perf}{ {\sf{perf}}  } 
\newcommand{\bef} { {\sf{bf}} } 
\newcommand{\st} {{\sf{st}}} 

\newcommand{\bX} { {\bf X} }
\newcommand{\ord}{{\mathcal{O}}}
\newcommand{\cb}{{\mathcal{C}}}
\newcommand{\cbt}{{\mathcal{ \widetilde{C} }}}
\newcommand{\littleo}{{\mathnormal{o}}}
\newcommand{\opt}{ {\sf opt} }
\newcommand{\bY}{ {\bf Y} } 

\newcommand{\gammarone} { \mu_{r,\hsppp 1}  } 
\newcommand{\gammartwo} {  \mu_{r, \hsppp 2} }
\newcommand{\gammatone} { {\mathrm{Gap}}_{t,\hsppp 12} }
\newcommand{\gammattwo} { {\mathrm{Gap}}_t }
\newcommand{\gammatc} { {\mathrm{Gap}}_t^c }
\newcommand{\gammarc} { \mu_{r, \hsppp 2}^{ c} }
\newcommand{\gammarcone}  {  \mu_{ r,\hsppp 2 }^{c, \hsppp ub}  }

\newcommand{\spatpow}{{\bf P}_{\sf spat}} 
\newcommand{\temppow}{ {\bf P}_{\sf temp} } 
\newcommand{\symbpow}{ {\bf P}_{\sf symb} } 

\newcommand{\bfx}{ \mathbf{x} } 
\newcommand{\bfw}{ \mathbf{w} } 
\newcommand{\bfa}{ \mathbf{a} } 
\newcommand{\bfc}{ \mathbf{c} } 
\newcommand{\thetah}{ \hat{\theta} } 
\newcommand{\thetab}{ \bar{\theta} } 
\newcommand{\etab}{ \bar{\eta} } 
\newcommand{\Reps}{ R_{\epsilon} } 
\newcommand{\pr}{ \textrm{Pr} } 
\newcommand{\magn}{ {\mathrm{Mag}} }
\newcommand{\magnf}{ {\mathrm{Mag}(\cdot)} }
\newcommand{\thetabm}{ {\boldsymbol{\theta}} }
\newcommand{\thetahbm}{ \hat{\boldsymbol{\theta}} }
\newcommand{\deltabm}{ {\boldsymbol{\delta}} }
\newcommand{\lambdabm}{ {\boldsymbol{\lambda}} }
\newcommand{\omegabm}{ {\boldsymbol{\omega}} }
\newcommand{\xibm}{ {\boldsymbol{\xi}} }

\begin{abstract} 
\noindent 
The focus of this work is on the analysis of transmit beamforming schemes with a low-rate feedback link in wireless sensor/relay networks, where nodes in the network need to implement beamforming in a distributed manner. Specifically, the problem of distributed phase alignment is considered, where neither the transmitters nor the receiver has perfect channel state information, but there is a low-rate feedback link from the receiver to the transmitters. In this setting, a framework is proposed for systematically analyzing the performance of distributed beamforming schemes. 
To illustrate the advantage of this framework, a simple adaptive distributed beamforming scheme that was recently proposed by Mudambai et al. is studied. Two important properties for the received signal magnitude function are derived. Using these properties and the systematic framework, it is shown that the adaptive distributed beamforming scheme converges both in probability and in mean. Furthermore, it is established that the time required for the adaptive scheme to converge in mean scales linearly with respect to the number of sensor/relay nodes.

\end{abstract}


\section{Introduction}

The problem of distributed beamforming arises quite naturally in wireless sensor/relay networks. In a sensor network, sensors make estimates of a common observed phenomenon and reach a consensus using a local message passing algorithm. In a relay network, a source node intends to communicate with the destination node by passing the message to all relay nodes. In both settings, the sensor/relay nodes then serve as distributed transmitters and seek to convey a common message to the intended receiver. To preserve energy in this stage, transmit beamforming has emerged as a promising scheme due to its potential array gain and low-complexity. 
However, perfect channel state information (CSI) at the transmitter is required by conventional transmit beamforming schemes to generate beamforming coefficients and achieve phase alignment at the receiver end. This requirement and the distributed nature of wireless sensor/relay networks make it difficult to implement transmit beamforming schemes in practice. 
Although obtaining perfect CSI may be too expensive from a practical point-of-view, partial CSI can be made available via a low-rate feedback link from the receiver to the transmitters.
As a consequence, there has been increased interest in designing efficient schemes that achieve distributed phase alignment in the presence of a low-rate feedback link~\cite{mudumbai_distrbf,mudumbai_thesis,thukral_distrbf_allerton_07,johnson_distrbf_isit08}. 
In this work, our goal is to provide a framework for systematically analyzing the performance of a general set of distributed beamforming schemes with such low-rate feedback. 

To illustrate the advantages of our framework, we focus on the analysis of a recently proposed training scheme for distributed beamforming~\cite{mudumbai_distrbf,mudumbai_thesis}. The proposed scheme is a simple adaptive algorithm using one bit of feedback information, and is attractive in practice since it is simple to implement. Naturally, one would expect a tradeoff in energy consumption due to possible slow convergence of distributed beamforming, but surprisingly, the scheme proposed in  \cite{mudumbai_distrbf} converges rapidly and hence utilizes energy efficiently. 
The scheme adjusts its phases for all sensors simultaneously in each time slot to achieve phase alignment. This reduces the overhead significantly compared with direct channel estimation between each source node and the destination node. In fact, the convergence time of the scheme scales linearly with the number of nodes. 

Although the scheme of   \cite{mudumbai_distrbf} has many desirable features, the fundamental reasons behind the effectiveness of the scheme are unclear from previous work.  In~\cite{mudumbai_thesis}, the analyses of the convergence and linear scalability of distributed beamforming schemes have been based on model approximations, which may be loose for some cases.
Assuming the stepsize approaches zero, stochastic approximation is used in~\cite{bucklew_2008} to show the convergence of the one-bit scheme in distribution. Furthermore, the authors proposed two more algorithms: the signed algorithm and the $\rho~\%$ solution algorithm and proved the convergence of both algorithms via the same technique. 
A discrete version of the problem has been solved in~\cite{thukral_distrbf_allerton_07,johnson_distrbf_isit08} by considering a simplified model with a binary channel and binary signaling. 

In this work, instead of focusing  on the convergence of a particular algorithm for a particular function, we seek a fundamental understanding into the convergence of distributed beamforming schemes more generally by studying them within the framework of local random search algorithms. Through this framework, we are able to provide a more comprehensive analysis of the fast convergence and linear scalability of the scheme proposed in \cite{mudumbai_distrbf}. In particular, our analysis does not involve approximation of any sort and hence 
makes statements on convergence and linear scalability in~\cite{mudumbai_thesis,bucklew_2008}
more rigorous. Our result is also stronger than that in~\cite{bucklew_2008} in the sense that convergence in probability is proved instead of convergence in distribution. Further, we show that due to the special structure of the objective function considered in this problem, \emph{any} adaptive distributed beamforming scheme that can be reformulated as a random search algorithm converges in probability. This broad set of algorithms also includes the signed and the $\rho~\%$ solution algorithms proposed in~\cite{bucklew_2008} and makes our analysis more general and rigorous than existing work in the literature.

We organize the paper as follows: 
In Section \ref{sec:system}, we introduce the system model and the received signal magnitude function, which is used as our metric to measure the beamforming array gain throughout the paper. In Section \ref{sec:adap_dist_bf}, we propose a framework that allows for a systematic analysis of a general set of adaptive distributed beamforming schemes. Specifically, we reformulate this set of adaptive distributed beamforming schemes as random search algorithms via a general framework. This reformulation provides insights into the necessary condition for the convergence of the scheme proposed in \cite{mudumbai_distrbf}. These insights lead us to investigate the properties of the received signal magnitude function in Section \ref{sec:converge_pf}. We further use these properties to prove the convergence of the local random search algorithm in probability and in mean, and provide simulations to validate our analysis. In Section \ref{sec:scale}, we  show that the time required for the algorithm to converge in mean scales linearly with the number of nodes. We also provide numerical results that validate our analysis. Finally, we conclude the paper in Section \ref{sec:conclude} and suggest directions for future research.

\section{System Setup}
\label{sec:system}

We consider the problem of distributed beamforming, where $n_s$ transmitters seek to beamform a common message to one receiver in a distributed manner.
We assume that each transmitter and the receiver is equipped with one antenna, and that the  channels from the transmitters to the receiver experience frequency-flat, slow fading. 
The discrete-time, complex baseband system model over a coherence interval is given by
\beq\label{eq:model}
y[t] = \sum_{i=1}^{n_s} h_i g_i[t] s[t] + w[t] = \sum_{i=1}^{n_s} a_i b_i[t] e^{j(\phi_i + \psi_i[t])} s[t] + w[t]
\eeq
where $s[t] \in \C$ is the transmitted common message, $y[t] \in \C$ is the received signal, and $w[t]\sim {\cal CN}(0,\sigma^2)$ corresponds to the additive white Gaussian noise. For transmitter $i$, we denote the channel fading gains by $h_i = a_i e^{j\phi_i}\in \C$ and beamforming coefficients by $g_i[t] = b_i[t] e^{j\psi_i[t]}\in \C$. Note that $a_i \geq 0$, $b_i[t] \geq 0$, and $\phi_i \in [0,2\pi]$, $\psi_i[t] \in [0,2\pi]$ for all $i$ and $t$ since they are the corresponding magnitudes and phases of $h_i$ and $g_i$, respectively. 
Moreover,  $a_i$ and $\phi_i$ are considered to be constant with time over the coherence interval  due to the slow fading assumption. We assume an average power constraint on $s[t]$ given by $E[|s[t]|^2] \leq P$ for all $t$.

We assume a noncoherent communication model, where the realization of the channel is unknown at both the transmitters and receiver. There is, however, an error-free, zero-delay feedback link of finite capacity from the receiver to all transmitters conveying low-rate partial channel state information (CSI) in each time step.

The goal of distributed beamforming is to pick the beamforming coefficients $\{g_i[t] = b_i[t] e^{j\psi_i[t]}\}$ to maximize the received $\snr$. 
In a noncoherent setting and with a low-rate feedback link, beamforming can only be achieved adaptively through training. 
Without loss of generality, we  assume that the signal $s[t]$ is constant during  the training stage. 
Furthermore, we make the following two simplifications. First, we assume that each transmitter utilizes the same amount of energy for each transmission, i.e., that $b_i[t] = 1$ for all $i$ and $t$, i.e., we do not optimize the beamforming gains, and we therefore set $s[t] = \sqrt{P}$. This assumption is justified for situations where the transmitters rely on a limited energy source (battery) and allowing them use different amounts of energy would cause some nodes to use up their energy before others. Secondly, we assume that the receiver can estimate the magnitude of the signal component\footnote{A good estimate of the received signal magnitude can be obtained directly when the noise is small, or by averaging over several time slots when the noise is not negligible.}  at the receiver (without the noise term $w[t]$ in \eqref{eq:model}). We therefore use received signal magnitude as the metric for optimizing the beamforming phases. 

The received signal magnitude can be expressed as
\beq\label{eq:mag}
\magn(\theta_1[t],\cdots,\theta_{n_s}[t]) = \sqrt{P} \left | \sum_{i=1}^{n_s} a_i e^{j\theta_i[t]} \right | 
\eeq
where $\theta_i[t] = \phi_i+\psi_i[t]$ is the total received phase for sensor $i$. 

It is easy to see that $\magnf$ is maximized when the phases $\{\theta_i[t]\}$ are aligned, i.e., they are equal to each other (modulo $2\pi$). Our goal is to study  {\em adaptive distributed beamforming schemes} that achieve this phase alignment through the use of a low-rate feedback link from the receiver.

\section{A Framework for Systematic Analyzing Adaptive Distributed Beamforming Schemes}
\label{sec:adap_dist_bf}

In this section, we introduce a framework for analyzing a general class of adaptive distributed beamforming schemes that can be reformulated as
random search algorithms. 
Random search algorithms are well studied in the literature~\cite{solis_min_rs_81,shi_np_gopt_00,tang_aprs_94} as methods to maximize an unknown function via random sampling. Once an adaptive distributed beamforming scheme can be successfully reformulated as a random search algorithm, 
a systematic study of the convergence of such adaptive scheme is possible. 

\subsection{Reformulation of Adaptive Distributed Beamforming Schemes as Random Search Algorithms}
\label{subsec:equivalence}

Adaptive distributed beamforming algorithms introduced in Section~\ref{sec:system} seek to maximize $\magnf$ given in \eqref{eq:mag} with the help of a low-rate feedback link. 
At each step of the adaptation, the signal magnitude at the receiver is a sample of the function  $\magnf$ . 
Thus, from the receiver point of view, the problem of distributed phase alignment can be considered under the setting of 
the following problem: 
\begin{problem}\label{prblm:un_max}
\emph{Given a unknown function $f: \Theta \rightarrow \R, \Theta \subseteq \R^n$, where only samples of $f(\thetabm)$ are available for arbitrary $\thetabm \in \Theta$, find the global maxima of $f$.}
\end{problem}

It is important to note that \emph{Problem \ref{prblm:un_max}} is a global maximization problem in general if no special structure is assumed for the objective function $f$. To solve the maximization in \emph{Problem \ref{prblm:un_max}}, one may be tempted to use gradient-based algorithms that are well-developed in the literature. Since it is possible for $f$ to possess local maxima, conventional gradient-ascent methods would fail in general. Besides, acquiring the gradient of the function $f$ may be infeasible especially when the function itself is unknown.
Hence, random search techniques~\cite{solis_min_rs_81,shi_np_gopt_00,tang_aprs_94} are more appropriate in this setting and can be described as follows: 

{\bf A Random Search Algorithm:}

\begin{itemize}

\item \emph{Step zero}: Initialize the algorithm by choosing $\thetabm[0] \in \Theta$.

\item \emph{Step one}: Generate a random perturbation $\deltabm[t]$ from the sample space $(\R^n, \mathcal{B}, \mu_t)$, where $\mathcal{B}$ is a Borel set on $\R^n$ and $\mu_t$ is a probability measure that could be time-varying. 

\item \emph{Step two}: Update the search point by $\thetabm[t] = D(\thetabm[t-1],\deltabm[t])$, where the map $D$ satisfies the condition $f(D(\thetabm[t-1],\deltabm[t])) \geq f(\thetabm[t-1])$.  

\end{itemize}

Clearly, for a random search algorithm, we require only function evaluations and  control over the probability measure $\mu_t$, which is used to sample the function. 
Any adaptive distributed beamforming scheme can be reformulated as 
a random search algorithm if each distributed transmitter initializes its phase as in \emph{Step zero}, generates a random perturbation of phase as in \emph{Step one}, and updates its new phase by the map $D$ as in \emph{Step two}. The low-rate feedback link is used to guarantee the condition $f(D(\thetabm[t-1],\deltabm[t])) \geq f(\thetabm[t-1])$. Note that the unknown function $f$ can be any objective function that we find fit for the distributed transmitters to optimize. This suggests that our framework can be used to analyze a more general function optimization problem over distributed networks. 
Note further that the probability measure $\mu_t$ for the sampling can be time-varying in general. The time-varying nature of the probability measure can be thought of as ``\emph{adaptive stepsize}'' for distributed algorithms in the most general sense. 
In this sense, our framework can be used to analyze a large set of adaptive distributed algorithms. 

\subsection{One-bit Adaptive Distributed Beamforming Scheme}
\label{subsec:describe_algm}

To illustrate the advantage of our framework, we now analyze a one-bit adaptive distributed beamforming scheme recently proposed in~\cite{mudumbai_distrbf}. Specifically, we reformulate this scheme as a local random search algorithm, which allows for its systematic analysis. We begin by describing the one-bit adaptive distributed beamforming scheme as follows:

{\bf A One-bit Adaptive Distributed Beamforming Scheme~\cite{mudumbai_distrbf}:}

\begin{itemize}

\item \emph{Step zero:} Referring to (\ref{eq:mag}) and noting that the $i$-th transmitter controls its beamforming phase $\psi_{i}[t]$, the algorithm is initialized by setting $\psi_{i}[0] = 0$, and hence $\theta_{i}[0] = \phi_i$ for transmitter $i$. 

\item \emph{Step one:} In this step, a random perturbation $\delta_i[t]$ is generated at each distributed transmitter such that $\{\delta_i[t]\}_{i=1}^{n_s}$ are i.i.d. uniform random variables in $[-\delta_0,\delta_0]$ across time and transmitters, where $\delta_0$ is a constant parameter. The random perturbation is added to the total phase of each transmitter. The distributed transmitters then use the perturbed total phases as their new total phases to transmit the training symbol. 

\item \emph{Step two:} After receiving the training symbols, the receiver measures the received signal magnitude and compares it with the signal magnitude received in the previous time slot. If the newly received signal magnitude is larger, the receiver feeds back a ``\emph{keep}'' beacon to the transmitters. Otherwise, a ``\emph{discard}'' beacon is sent to the transmitters. Note that the beacon is a broadcast from the receiver to all transmitters. Clearly, this feedback scheme only requires one bit of feedback information per time step. 
When a ``\emph{keep}'' is received at the transmitters, each transmitter selects and keeps its newly updated total phase. Otherwise, the old phase is selected and the new phase discarded. This selection process is determined by whether the random perturbation increases or decreases the array gain for the adaptive distributed beamforming scheme. Specifically, the evolution of $\thetabm[t]$ is given by
\beq\label{eq:evol_theta}
\thetabm[t] = \left\{ \begin{array}{ll}
\thetabm[t-1] + \deltabm[t], & \mathrm{if} \hsp \deltabm[t] \in \mathcal{K}\\
\thetabm[t-1], & \mathrm{if} \hsp \deltabm[t] \notin \mathcal{K}
\end{array}
\right.
\eeq
where $\thetabm[t] = [\theta_1[t],\cdots,\theta_{n_s}[t]]^T$, $\deltabm[t] = [\delta_1[t],\cdots,\delta_{n_s}[t]]^T$, and $\mathcal{K} = \{\deltabm[t] \| \magn(\thetabm[t-1] + \deltabm[t]) > \magn(\thetabm[t-1])\}$. 
\end{itemize}

Matching the steps of the above one-bit adaptive scheme and those of a random search algorithm introduced in Section~\ref{subsec:equivalence}, it is clear that the one-bit adaptive distributed beamforming algorithm can be regarded as a special case of the random search algorithm by setting 
\beqa \label{eq:mu}
f &=& \magnf \\
n &=& n_s \\ 
\Theta &=& [0,2\pi]^{n_{s}} \\ \label{eq:theta_set}
\mu_t &=& \mu \\ \label{eq:d_fcn}
D(\thetabm[t-1],\deltabm[t]) &=& 
\thetabm[t-1]
+1_{\left\{\deltabm[t] \in \mathcal{K} \right\}}\deltabm[t]
\eeqa
where $1_{\{\cdot\}}$ is the indicator function and $\mu$ is uniform on $[-\delta_0, \delta_0]^{n_s}$, which is a $n_s$-dimensional hypercube. Note that (\ref{eq:d_fcn}) is the same as the evolution described by (\ref{eq:evol_theta}). 

Since the probability measure $\mu$ is non-zero only within a hypercube, with sides of length $2\delta_0$ and centered around $\thetabm[t-1]$, the one-bit adaptive distributed beamforming scheme can be reformulated as a \emph{local} random search algorithm. We emphasize again that we can use this framework to study more general adaptive distributed beamforming schemes. For example, the probability measure for sampling may be time-varying and with a support that spans the entire space $\Theta$. We can also study adaptive distributed beamforming schemes with more than one bit of feedback information. It is also interesting to note the connection between this local random search algorithm and simulated annealing~\cite{kirkpatrick_sa}.  Simulated annealing is a generic probabilistic algorithm that approximates the global optimal solution of a given function in a large search space. The algorithm uses a parameter $T$ called the \emph{temperature} to control the acceptance probability, i.e., the probability that the current state of the algorithm transitions to a new state. If we let $T\rightarrow 0$ and assume that the current state is only allowed to move to neighboring states, the simulated annealing procedure reduces to a local random search algorithm.

A local random search algorithm, however, does not necessarily converge in general. For example, if the unknown function possesses local maxima (that are not global maxima), the sequence $\left\{ \thetabm[t] \right\}_{t=0}^\infty$ is likely to be trapped in a local maximum  if the local perturbation $\delta_0$ is not large enough.  
Thus, a necessary condition for the convergence of local random search algorithms for arbitrary $\delta_{0}$ is that there is no local maximum point for $\magnf$. 
With these in mind, two questions arise naturally: $a)$ Does the reformulated local random search algorithm even converge? $b)$ If it does, is there a fundamental reason behind the convergence? In the following section, we investigate properties of the function $\magnf$ towards the goal of addressing these questions.

\section{Convergence of the Distributed Beamforming Scheme}
\label{sec:converge_pf}

\subsection{Properties of Received Signal Magnitude Function}
\label{sec:property}

The properties of the received signal magnitude function $\magnf$ do not depend on the time evolution of its arguments. We hence ignore the time dependence of $\theta[t]$ in this section. The following proposition states the first property of $\magnf$. 

\begin{prop}\label{prop:local_global_max}
For the received signal magnitude function $\magnf$ defined in (\ref{eq:mag}), all local maxima are global maxima.
\end{prop}

\begin{proof}
To facilitate analysis, we introduce a change of variables 
\beq\nonumber
\bfx_i := \left[ \begin{array}{c}
x_i^R \\
x_i^I
\end{array}
\right] 
= \left[ \begin{array}{c}
\cos\theta_i \\
\sin\theta_i
\end{array}
\right] 
\eeq
Eqn. (\ref{eq:mag}) can be rewritten as
\beq\nonumber
\magn (\bfx_1,\cdots, \bfx_{ns}) = \sqrt{P} \left\| \sum_{i=1}^{n_s} a_i \bfx_i \right\| 
\eeq
where $\|\bfx_i\|^2 = 1$ for all $i=1,\cdots,n_s$.
The maximization of $\magnf$ can be rewritten as
\beq\label{eq:opt}
\max_{\|\bfx_i\|^2 = 1, i=1,\cdots,n_s} \left\| \sum_{i=1}^{n_s} a_i \bfx_i \right\|^2
\eeq
In the following, we will show that all local maxima of this objective function correspond to complete phase alignment for all transmitters. That is, all local maximum points are global maximum points. 

By relaxing the equality constraints to inequality constraints, the optimization problem in (\ref{eq:opt}) is equivalent to  
\beq\label{eq:opt_2}
\max_{\|\bfx_i\|^2 \leq 1, i=1,\cdots,n_s} \left\| \sum_{i=1}^{n_s} a_i \bfx_i \right\|^2
\eeq
This equivalence can be seen as follows: if $\bfx^*$ is a local maximum with an inactive constraint $\|\bfx_k^*\|^2<1$, by fixing all other variables $\{\bfx_j^*\}_{j\neq k}$, we obtain
\beq\nonumber
\left\| \sum_{i=1}^{n_s} a_i \bfx_i^* \right\|^2 = \left\| a_k \bfx_k^* + \bfc \right\|^2 
= ( a_k {x_k^*}^R + c^R )^2 + (a_k {x_k^*}^I + c^I )^2
\eeq
where $\bfc = [c^R c^I]^T$ is a constant vector depending on $\{\bfx_j^*\}_{j\neq k}$. 
Obviously, the above function can be improved by appropriately perturbing $\|\bfx_k^*\|$ according to the signs of $c^R$ and $c^I$. This contradicts the fact that $\bfx^*$ is a maximum. Thus, all constraints are active if $\bfx^*$ is a maximum point. This shows that the optimization problems (\ref{eq:opt}) and (\ref{eq:opt_2}) are equivalent. 

Focusing on the optimization problem with relaxed constraints, the Lagrangian of (\ref{eq:opt_2}) reads
\beq\nonumber
L(\bfx,\lambdabm) =  -\|\bfw\|^2 + \sum_{i=1}^{n_s} \lambda_i (\|\bfx_i\|^2-1)
\eeq
where $\bfx = [\bfx_1^T,\cdots,\bfx_{n_s}^T]^T$, $\lambdabm = [\lambda_1,\cdots,\lambda_{n_s}]^T$, $\lambda_i \geq 0$ for all $i=1,\cdots,n_s$, and $\bfw = \sum_{i=1}^{n_s} a_i \bfx_i$. By the Lagrange Multiplier Theorem, all local maxima satisfy  
\beqa\label{eq:1storder_cond}
\nabla_{\bfx_i} L(\bfx,\lambdabm) &=& -2 a_i \bfw^T + 2 \lambda_i \bfx_i^T = \mathbf{0}^T\\
\sum_{i=1}^{n_s}\lambda_i (\|\bfx_i\|^2-1) &=& 0\\
\|\bfx_i\|^2-1&\leq& 0 
\eeqa
for all $i = 1,\cdots,n_s $.
Let $\bfx^*$ be a local maximum and $\lambdabm^*$ be the corresponding Lagrange multipliers. If $\lambda_i^* = 0$, Eqn. (\ref{eq:1storder_cond}) implies that $\bfw = \mathbf{0}$ since\footnote{Note that the case where $a_i = 0$ is not interesting since we can always reduce the dimension of the problem by ignoring $\bfx_i$} $a_i>0$. In this case, $\magn\left(\bfx^*\right) = 0$ and this contradicts the fact that $\bfx^*$ is a local maximum, since we can always improve $\magnf$ by letting $\bfx_i^* = [\xi \hspp 0]^T$, $\xi \leq 1$, and $\bfx_j = \mathbf{0}$ for all $j\neq i$. This leads to $\lambda_i>0$ for all $i$. We hence have
\beqa\label{eq:x_opt}
\bfx_i^* &=& \frac{a_i}{\lambda_i^*} \bfw \\ \label{eq:lambda_opt}
\lambda_i^* &=& a_i \|\bfw\|
\eeqa

The optimal solutions described by (\ref{eq:x_opt}) and (\ref{eq:lambda_opt}), however, also satisfy
\beq\nonumber
\magn (\bfx^*) = \sqrt{P}\left\| \sum_{i=1}^{n_s} a_i \frac{\bfw}{\|\bfw\|} \right\| = \sqrt{P}\sum_{i=1}^{n_s} a_i
\eeq
and hence are global maxima. This completes our proof. 

\end{proof}

\emph{Proposition}~\ref{prop:local_global_max} implies that the local random search algorithm cannot be trapped in a suboptimal local maximum since all local maxima are global maxima. 
Furthermore, it also suggests that the necessary condition for the convergence of random search algorithms is satisfied. While it is intuitively clear that the local random search algorithm should converge according to \emph{Proposition}~\ref{prop:local_global_max}, it is to be noted that the condition is only necessary and may not be sufficient. 
We will provide a rigorous proof of the convergence of the local random search algorithm later. Now, we explore an additional property of $\magnf$ that explains the efficiency of the algorithm. 

Another interesting property of $\magnf$ is that it is invariant under a common phase shift to all transmitters. That is,
\beqa\nonumber
\magn(\thetabm+\theta_c \mathbf{e}) &=& \sqrt{P} \left| \sum_{i=1}^{n_s} a_i e^{j\left(\theta_i+\theta_c\right)} \right| \\ \nonumber
&=& \sqrt{P} \left| e^{j\theta_c}\sum_{i=1}^{n_s} a_i e^{j\theta_i} \right| = \magn(\thetabm)
\eeqa
where $\mathbf{e}$ is a $n_s \times 1$ vector with all elements equal to one, and $\theta_c$ is a common phase shift that can depend on $\{\theta_i\}_{i=1}^{n_s}$. One possible choice for the common phase shift is to let $\theta_c(\theta_1,\cdots,\theta_{n_s})$ be such that the imaginary part within the modulus function is canceled, i.e.,
\beqa\nonumber
\magn(\thetabm) &=& \magn(\thetabm+\theta_c(\theta_1,\cdots,\theta_{n_s}) \mathbf{e}) \\\nonumber
&=& \sqrt{P}\sum_{i=1}^{n_s} a_i \cos\left(\theta_i+\theta_c(\theta_1,\cdots,\theta_{n_s})\right) \\\nonumber
&=& \sqrt{P}\sum_{i=1}^{n_s} a_i \cos\theta_i' = \magn(\thetabm')
\eeqa
where $\thetabm' = [\theta_1',\cdots,\theta_{n_s}']^T$.
Note that in the shifted $\thetabm'$ domain, the global maxima occur only when $\theta'_i = 0$ or $2k\pi$ for all $i$, where $k$ is any integer. The shift-invariant property results in multiple global maxima for the function $\magnf$. In fact, all global maxima form a one-dimensional ``ridge'' since if $\thetabm^*$ is a global maximum, $\bar{\boldsymbol{\theta}}$ with $\thetab_i = \theta^*_i + \theta_c$ is also a global maximum. This property leads to the rapid convergence of the local random search algorithm since converging to any of these global maximum points is adequate. 

We conclude this section by summarizing these two important properties of $\magnf$ as follows:

\begin{enumerate}
\item all local maxima are global maxima, and 
\item a common shift to its arguments does not change its value.
\end{enumerate}

\subsection{Proof of Convergence}
\label{sec:convg_proof}

Intuitively, \emph{Property} 1 guarantees the convergence of any local random search algorithm. To make this precise, we introduce an $\epsilon$-convergence region
\beq\label{eq:def_Reps}
\Reps = \left\{ \thetabm \in \Theta : \magn(\thetabm) > \magn\left(\thetabm^{*}\right) -\epsilon \right\}
\eeq
where $\thetabm^{*}$ is the optimal total phase and satisfies $\magn(\thetabm^{*}) = \sqrt{P}\sum_{i=1}^{n_{s}} a_{i}$. We define the convergence of a random search algorithm in probability as follows:
\begin{defn}
A sequence $\left\{\thetabm[t]\right\}_{t=0}^{\infty}$ generated by a random search algorithm is said to be convergent in probability if, given $\epsilon>0$,
\beq\nonumber
\lim_{t \rightarrow \infty} \pr\left[ \thetabm[t] \in \Reps \right] = 1
\eeq
In other words, $\magn(\thetabm[t])$ converges to $\magn(\thetabm^*)$ in probability.
\end{defn}

For the proof of convergence, we futher derive a proposition stating that for any $\thetabm$ outside of $\Reps$, there is a non-zero probability of improving $\magnf$ by applying a local perturbation to $\thetabm$. 

\begin{prop}\label{prop:pos_improve}
For any given $\thetabm \in \Theta \setminus \Reps$ and $\delta_{0} > 0$, there correspond $\gamma >0$ and $0< \eta \leq 1$ such that
\beq\nonumber
\pr \left[ \magn(\thetabm+\deltabm) - \magn(\thetabm) \geq \gamma \right] \geq \eta
\eeq
where $\deltabm$ is a random vector with i.i.d. elements uniformly distributed over $[-\delta_{0}, \delta_{0}]$.
\end{prop}
\begin{proof}
From \emph{Proposition \ref{prop:local_global_max}}, all local maxima are global maxima for the function $\magnf$. This implies that for all $\thetabm \notin \Reps$ and all $\delta_{0}>0$, there exists a point $\thetabm_{u} \in S_{\thetabm}$ and a constant $\gamma(\thetabm)>0$ such that
\beq\label{eq:localnomax}
\magn(\thetabm_{u}) - \magn(\thetabm) \geq 2 \gamma(\thetabm)
\eeq
where the set $S_{\thetabm}$ is a hypercube of length $2\delta_{0}$ centered around $\thetabm$ given by
\beq\nonumber
S_{\thetabm} = \left\{ \omegabm \in \Theta : \omegabm = \thetabm+\deltabm, \hspp \deltabm \in [-\delta_{0}, \delta_{0}]^{n_{s}} \right\}
\eeq
The continuity of $\magnf$ implies that there exists $\sigma(\thetabm_{u}) >0$ such that for all $\xibm \in T := \left\{ \omegabm \in \Theta : \| \omegabm \| \leq \sigma(\thetabm_{u})\right\}$, we have
\beq\label{eq:cont}
\left| \magn(\thetabm_{u}+\xibm) - \magn(\thetabm_{u})\right| \leq \gamma(\thetabm) 
\eeq
Combining (\ref{eq:localnomax}) and (\ref{eq:cont}), we arrive at a lower bound
\beqa\nonumber
\magn(\thetabm_{u}+\xibm) - \magn(\thetabm)  &=& \magn(\thetabm_{u}+\xibm) - \magn(\thetabm_{u}) \\\nonumber
&+& \magn(\thetabm_{u}) - \magn(\thetabm) \\\nonumber
&\geq& -\gamma(\thetabm) + 2 \gamma(\thetabm) = \gamma(\thetabm)
\eeqa
Referring to (\ref{eq:mu}) for the definition of $\mu$, the above lower bound leads to 
\beq\nonumber
\pr \left[ \magn(\thetabm+\deltabm) - \magn(\thetabm) \geq \gamma(\thetabm) \right] \geq \mu \left( T \right) =: \eta(\thetabm)
\eeq
Note that $\mu(T)$ is a function of $\thetabm$, since $\thetabm_{u}$ is a function of $\thetabm$. We complete the proof of the proposition by letting 
\beqa\nonumber
\gamma &=& \inf_{\thetabm \in \Theta \setminus \Reps} \gamma(\thetabm) \\\nonumber
\eta &=& \inf_{\thetabm \in \Theta \setminus \Reps} \eta(\thetabm)
\eeqa
\end{proof}

Note that the proof of this proposition can easily be generalized for any local random perturbation $\deltabm$.
Since before the sequence reaches the $\epsilon$-convergence region, there is always a non-zero probability of improving $\magnf$ for each time step, the convergence of the sequence is to be expected. A simple deterministic analogue is the convergence of a monotonically non-decreasing function. The probabilistic nature of the algorithm complicates the proof. This will become clear in the proof of our next theorem. 

\begin{thm}\label{thm:converge}
For the function $\magnf$ defined in (\ref{eq:mag}), let $\left\{\thetabm[t]\right\}_{t=1}^\infty$ be a sequence generated by the local random search algorithm described in Eqn. (\ref{eq:mu})-(\ref{eq:d_fcn}). Then the resulting sequence converges in probability, i.e., given $\epsilon>0$,
\beq\nonumber
\lim_{t \rightarrow \infty} \pr\left[ \thetabm[t] \in \Reps \right] = 1
\eeq
\end{thm}
\begin{proof}
By \emph{Proposition} \ref{prop:pos_improve}, we know that given any time $t$
\beq\nonumber
\pr \left[ \left\{ \magn(\thetabm[t-1]+\deltabm[t]) - \magn(\thetabm[t-1]) \geq \gamma \right\} \hspp \textrm{or} \hspp \left\{ \thetabm \in \Reps \right\} \right] \geq \etab
\eeq
where $\etab = \min \left\{ \pr[\thetabm \in \Reps], \eta \right\}$. Since $\Theta$ is compact and $\magnf$ is continuous, there always exists a positive integer $p$ such that
\beq\nonumber
p\gamma > \magn(\thetabm_{1}) - \magn(\thetabm_{2}), \hspp \forall \thetabm_{1}, \thetabm_{2} \in \Theta
\eeq
The probability that the sequence lies in $\Reps$ after $p$ time steps is hence lower bounded by
\beq\nonumber
\pr \left[ \thetabm[p] \in \Reps \right] \geq \etab^{p}
\eeq
since $\{\deltabm[t]\}_{t=0}^\infty$ are independent across time.
This leads to $\pr \left[ \thetabm[p] \notin \Reps \right] \leq 1-\etab^{p}$ and 
\beq\nonumber
\pr \left[ \thetabm[pm] \in \Reps \right]  = 1-\pr \left[ \thetabm[pm] \notin \Reps \right] \geq1-(1- \etab^{p})^{m}
\eeq
for all $m=1, 2,\cdots$.
The lower bound is still valid if we let the sequence progress $\ell$ time steps further, i.e., 
\beq\nonumber
\pr \left[ \thetabm[pm+\ell] \in \Reps \right] \geq1-(1- \etab^{p})^{m}
\eeq
for all $m = 1, 2,\cdots, \ell = 0,\cdots, p-1$.
We complete the proof by letting $m \rightarrow \infty$.

\end{proof}

\emph{Theorem \ref{thm:converge}} states that the local random search algorithm in  (\ref{eq:mu})-(\ref{eq:d_fcn}) converges in probability, and hence also provides a proof of convergence for the one-bit adaptive distributed beamforming scheme in (\ref{eq:evol_theta}). In particular, \emph{Theorem \ref{thm:converge}}  implies the convergence of the sequence $\{\magn(\thetabm[t])\}_{t=0}^\infty$ in probability. Since the sequence is non-negative and monotonically non-decreasing, we can conclude that $\{\magn(\thetabm[t])\}_{t=0}^\infty$ also converges in mean by the Monotone Convergence Theorem~\cite{durrett}. Further, by properly generalizing \emph{Proposition} \ref{prop:pos_improve}, it is straightforward to show that \emph{any} adaptive 
distributed beamforming scheme that can be reformulated a local random search algorithm and seeks to maximize 
\emph{any} objective function that satisfies \emph{Property} $1$ converges in probability.

\begin{figure}[htb!]
\centering
\includegraphics[height=2.5in,width=3.5in]{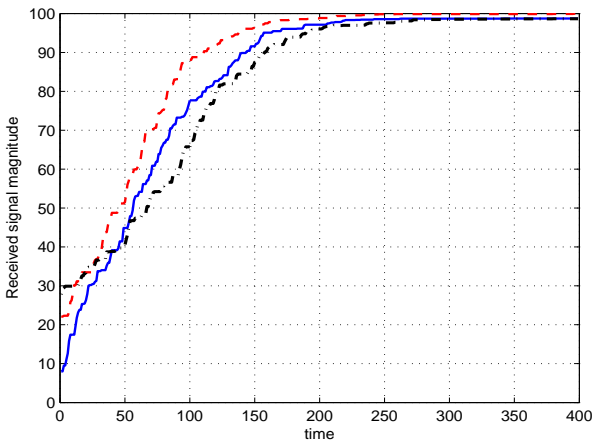}
\caption{Evolutions of sequences generated by the adaptive distributed beamforming scheme.}\label{fig:sample_path}
\end{figure}

In Fig.~\ref{fig:sample_path}, we illustrate the evolution of the sequences generated by the local random search algorithm from different initial points. The initial points are generated randomly from a uniform distribution over $\Theta$. Only three sample paths of the sequence are included in the figure since similar behaviors can be observed for other sample paths. 
For each iteration, the random perturbation $\delta_{i}$ for the $i$th transmitter is a uniform random variable over $[-\delta_{0}, \delta_{0}]$, where $\delta_{0} = \pi/30$. Note that we use the same channel coefficients to generate these sequences since the focus here is on the effect of different initial points. In particular, the channel coefficients are randomly generated from i.i.d. $\mathcal{CN}(0,1)$ in the beginning of the simulation, and remain fixed afterwards. 

From the figure, we observe the rapid convergence of the local random search algorithm, irrespective of where it is initialized.
We emphasize again that the fast convergence results follow from the two important properties for the function $\magnf$ as discussed in Section \ref{sec:property}. \emph{Property} $1$ guarantees the convergence of the local search algorithm; \emph{Property} $2$ results in multiple global maxima for the function $\magnf$ and hence the fast convergence of the algorithm. 
The simulations provide a partial validation of our proof since we would expect the convergence to fail from some initial points if there were non-optimal local maxima for $\magnf$. 
It is to be noted that the convergence of the local random search algorithm does not guarantee that it is the most efficient scheme in terms of the number of function evaluations, and hence the most efficient scheme in terms of energy. However, the algorithm does have a desirable scaling property, i.e., the time required for the algorithm to converge in mean scales linearly with the number of transmitters. This is the topic of the following section.

\section{Scaling Law}
\label{sec:scale}

Due to the probabilistic nature of the local random search algorithm, we defined convergence in probability in Section \ref{sec:convg_proof} and showed that the local random search algorithm converges. For the analysis of the scaling law, however, we can only show convergence in mean, which is defined as follows:
\begin{defn}
A sequence $\{\thetabm[t]\}$ generated by a random search algorithm is said to converge in mean if there exists $t_N \geq 0$ such that 
\beq\nonumber
E_{\{\deltabm[\tau]\}_{\tau=0}^{t} | \bfa, \thetabm[0]}\left[\magn\left( \thetabm[t] \right)\right]> \magn\left( \thetabm^* \right) - \epsilon = \sqrt{P}\sum_{i=1}^{n_s} a_i - \epsilon
\eeq
for all $t\geq t_N$, where $\bfa = [a_1,\cdots,a_{n_s}]^T$. That is, $\magn(\thetabm[t])$ converges to $\magn(\thetabm^*)$ in mean. 
\end{defn}

In this section, our goal is to find the time required for the local random search algorithm to converge in mean, starting from any initial point. In other words, we are interested in finding the \emph{hitting time}\footnote{The hitting time in this work is defined as the time required for the algorithm to converge in mean.
} of the random search algorithm, and determining its behavior as a function of the number of transmitters. 
Specifically, we derive an upper bound on the hitting time of the local random search algorithm as a function of $n_s$.  
Note that the study of the hitting time makes sense only if the sequence indeed converges in mean, which we established in Section \ref{sec:convg_proof}.

To facilitate analysis, we define the increment function of $\magnf$ at time $\tau$ as
\beqa \nonumber
I[\tau] &=& \left[\magn\left(\thetabm[\tau]\right)-\magn\left(\thetabm[\tau-1]\right)\right]^+ \\\label{eq:increment}
&=& \left[ \magn(\thetabm[\tau-1]+\deltabm[\tau]) - \magn(\thetabm[\tau-1] \right]^+
\eeqa
where $[x]^+ = \max(x,0)$. We then rewrite the received signal magnitude function at any given time $k_0 n_s$ as
\beq\label{eq:mag_equivalent}
\magn\left(\thetabm[k_0 n_s]\right) = \sum_{\tau=1}^{k_0 n_s} I[\tau] + \magn\left(\thetabm[0]\right) =: \sum_{\tau=1}^{k_0 n_s} I[\tau] + c_0
\eeq
where $k_0$ is a positive integer and $c_0 \geq 0$.

From \emph{Proposition \ref{prop:pos_improve}} we have that for any given $\tau$ such that $\thetabm[\tau-1] \notin \Reps$ and any local random perturbation $\deltabm[\tau]$, there correspond $\gamma >0$ and $0< \eta \leq 1$ such that
\beq\nonumber
\pr \left[ \magn(\thetabm[\tau-1]+\deltabm[\tau]) - \magn(\thetabm[\tau-1]) \geq \gamma \right] \geq \eta
\eeq
Thus, we have 
\beqa 
\begin{split} \nonumber
&E_{\deltabm[\tau] | \bfa, \thetabm[\tau-1]}\left[ I[\tau] \right] \\
&\geq \gamma \pr \left[ \magn(\thetabm[\tau-1]+\deltabm[\tau]) - \magn(\thetabm[\tau-1]) \geq \gamma \right] \\
&\geq \gamma\eta > 0
\end{split}
\eeqa
for any $\tau$ such that $\thetabm[\tau-1] \notin \Reps$. Referring to (\ref{eq:increment})-(\ref{eq:mag_equivalent}), we obtain
\beqa 
\begin{split} \nonumber
&E_{\{\deltabm[\tau]\}_{\tau=0}^{k_0 n_s} | \bfa, \thetabm[0]}\left[\magn\left(\thetabm[k_0 n_s]\right)\right] \\
&=\sum_{\tau=1}^{k_0 n_s} E_{\deltabm[\tau] | \bfa, \thetabm[\tau-1]}\left[I[\tau]\right] + c_0 \geq k_0 n_s \gamma\eta+c_0 \geq \sqrt{P}\sum_{i=1}^{n_s} a_i 
\end{split}
\eeqa
where the last inequality follows by choosing $k_0 = \left\lceil \frac{\sqrt{P}\max_i\{a_i\}}{\gamma\eta}\right\rceil$.
This implies that the hitting time for the local random search algorithm is at most $k_0n_s$, from any initial point. Hence, the hitting time for the algorithm scales linearly with the number of transmitters.

\begin{figure}[htb!]
\centering
\includegraphics[height=2.5in,width=3.5in]{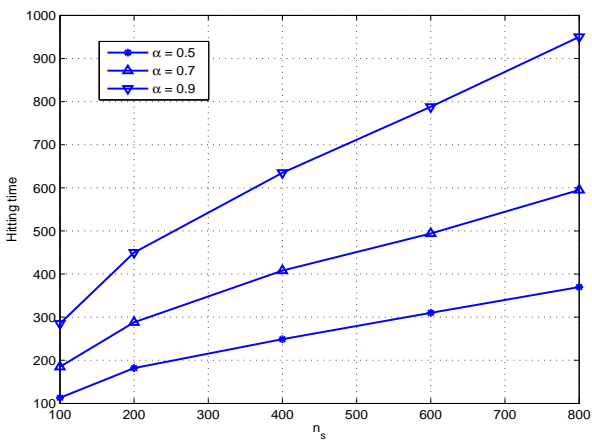}
\caption{Hitting time for the adaptive distributed beamforming scheme with different values of $\alpha$.}\label{fig:hitting_time}
\end{figure}

\begin{figure}[htb!]
\centering
\includegraphics[height=2.5in,width=3.5in]{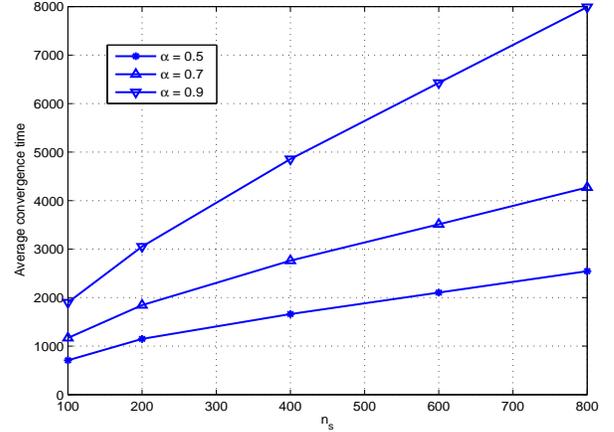}
\caption{Average convergence time for the adaptive distributed beamforming scheme with different values of $\alpha$.}\label{fig:ave_cge_time}
\end{figure}

In our simulations, we say that the sequence converges to the $\alpha$ fraction of the global maxima if $\magn(\thetabm[t])\geq \alpha \magn(\thetabm^*)$. We assume that channel coefficients are i.i.d. complex Gaussian variables $\mathcal{CN}(0,1)$, and use the origin as our initial point. We set $\delta_0 =\pi/90 $ for all our simulations. Fig.~\ref{fig:hitting_time} demonstrates the hitting time required for the adaptive distributed beamforming scheme to converge in a relative sense when $\alpha = 0.5,$ $0.7,$ and $0.9$. 
It is clear that the hitting time increases as $\alpha$ increases. The scaling law for the hitting time with respect to $n_{s}$, however, is the same for all values of $\alpha$. Indeed, 
we observe linear scaling for all values of $\alpha$. This observation confirms our theoretical analysis.
Fig.~\ref{fig:ave_cge_time} shows the average convergence time for the adaptive distributed beamforming scheme to within a fraction of the globally maximum value $\alpha\magn(\thetabm^*)$, for different values of $\alpha$. It is important to note the difference between the hitting time and the average convergence time.
Since our algorithm is probabilistic in nature, the convergence time is essentially a random variable and each run of the algorithm provides a sample  for this random variable. 
Fixing the number of transmitters $n_s$, we obtain the average convergence time by averaging over a hundred samples of this random variable, while the hitting time is obtained by comparing $E[\magn(\thetabm[t])]$ with $\alpha\magn(\thetabm^*)$. From Fig.~\ref{fig:ave_cge_time}, we observe the same linear scaling behavior for the average convergence time. We expect this property for the average convergence time can be shown in a similar manner.

\section{Concluding Remarks and Future Work}
\label{sec:conclude}

In this work, we have proposed a framework that allows for a systematic analysis of adaptive distributed beamforming schemes in sensor/relay networks. We used this framework to study the convergence and scaling law of a recently proposed one-bit adaptive distributed beamforming scheme~\cite{mudumbai_distrbf}. We first reformulated the one-bit adaptive scheme as a local random search algorithm. This reformulation provided insights into the convergence of the one-bit adaptive scheme, and led us to investigate the fundamental properties for the received signal magnitude function $\magnf$. We identified two important properties of the function that contribute to the rapid convergence of the algorithm. First, all local maxima are global maxima. This prevents any local random search algorithm from being trapped in non-optimal local maximum points. Secondly, the $\magnf$ function is invariant under a common shift to its arguments. This property results in multiple global maximum points for $\magnf$ and hence the rapid convergence of the algorithm. Based on these properties, we have shown the convergence of the algorithm, both in probability and in mean.
We further provided an upper bound on the hitting time of the algorithm, and demonstrated that the hitting time scales linearly with the number of sensor/relay nodes. This linear scaling is desirable, especially when the network is densely populated. We have also provided simulations that validate our analysis. 

It is important to note that the effectiveness of the one-bit adaptive distributed beamforming scheme depends critically on the properties of the function $\magnf$. 
Maximizing $\magnf$ is equivalent to maximizing the received $\snr$ if there is no error in obtaining the common message, which is true in the training stage since the common message is simply fixed and known to the receiver. 
On the other hand if adaptation is being performed blindly (without training) it would be necessary to consider the possibility of errors in common message. The corresponding objective function may then  not possess the same desirable properties as $\magnf$, e.g., the objective function may possess  local maxima that are not global maxima. Much work needs to be done to understand how our results can be applied in this more complicated scenario. One thing that is clear, however, is that we will need to develop new algorithms that exploit the global structure of the new objective function since local algorithms can be trapped in local maxima. Our general framework for studying adaptive beamforming algorithms is even more useful in this context since it connects the problem to a well-studied field of global optimization algorithms.

\bibliographystyle{ieeetr}
\bibliography{newrefs}

\begin{biography}
{Che Lin (S'02--M'08)} received the B.S. degree in Electrical Engineering from National Taiwan University, Taipei, Taiwan, in 1999. He received the M.S. degree in Electrical and Computer Engineering in 2003, the M.S. degree in Math in 2008, and the Ph.D. degree in Electrical and Computer Engineering in 2008, all from the University of Illinois at Urbana-Champaign, IL. Since 2008, he has been at National Tsing Hua University, where he is currently an assistant professor. 
    Dr. Lin received a two-year Vodafone graduate fellowship in 2006, the E. A. Reid fellowship award in 2008, and holds a U.S. patent, which has been included in the 3GPP LTE standard.
    His research interests include feedback systems, distributed algorithms in networks, practical MIMO code design in general networks, optimization theory, and information theory.
\end{biography}

\begin{biography}
{Venugopal V. Veeravalli (S'86--M'92--SM'98--F'06)} received the Ph.D.
degree in 1992 from the University of Illinois at Urbana-Champaign, the
M.S. degree in 1987 from Carnegie-Mellon University, Pittsburgh, PA,
and the B.Tech degree in 1985 from the Indian Institute of Technology,
Bombay, (Silver Medal Honors), all in Electrical Engineering. He
joined the University of Illinois at Urbana-Champaign in 2000, where
he is currently a Professor in the department of Electrical and
Computer Engineering, and a Research Professor in the Coordinated
Science Laboratory. He served as a program director for communications
research at the U.S. National Science Foundation in Arlington, VA
from 2003 to 2005. He has previously held academic positions at
Harvard University, Rice University, and Cornell University.

His research interests include distributed sensor systems and
networks, wireless communications, detection and estimation theory,
and information theory. He is a Fellow of the IEEE and was on the
Board of Governors of the IEEE Information Theory Society from 2004 to
2007. He was an Associate Editor for Detection and Estimation for the
IEEE Transactions on Information Theory from 2000 to 2003, and an
associate editor for the IEEE Transactions on Wireless Communications
from 1999 to 2000. Among the awards he has received for research and
teaching are the IEEE Browder J. Thompson Best Paper Award, the
National Science Foundation CAREER Award, and the Presidential Early
Career Award for Scientists and Engineers (PECASE). He is a
distinguished Lecturer for the IEEE Signal Processing Society for
2010-2011.
\end{biography}

\begin{biography}
{Sean P. Meyn}
received the B.A. degree in Mathematics Summa Cum Laude from UCLA in 1982, and the PhD degree in Electrical Engineering from McGill University in 1987. After a two year postdoctoral fellowship at the Australian National University in Canberra, Dr. Meyn and his family moved to the Midwest. He is now a Professor in the Department of Electrical and Computer Engineering, and a Research Professor in the Coordinated Science Laboratory at the University of Illinois, where he is director of the Decision and Control Lab.. He is an IEEE fellow.

He is coauthor with Richard Tweedie of the monograph Markov Chains and Stochastic Stability, Springer-Verlag, London, 1993,  and received jointly with Tweedie the 1994 ORSA/TIMS Best Publication In Applied Probability Award.  The 2009 edition is published in the Cambridge Mathematical Library.  His new book, Control Techniques for Complex Networks is published by Cambridge University Press.

He has held visiting positions at universities all over the world, including the Indian Institute of Science, Bangalore during 1997-1998 where he was a Fulbright Research Scholar. During his latest sabbatical during the 2006-2007 academic year he was a visiting professor at MIT and United Technologies Research Center (UTRC). His research interests include stochastic processes, control and optimization, complex networks, and information theory. Current funding is provided by NSF, Dept. of Energy, AFOSR, and DARPA. 
\end{biography}

\end{document}